\documentclass[%
 reprint,
 superscriptaddress,
 amsmath,amssymb,
 aps,
 prc,
 floatfix,
 showkeys
]{revtex4-1}

\usepackage{graphicx}
\usepackage{dcolumn}
\usepackage{bm}
\usepackage{hyperref}
\usepackage[mathlines,modulo]{lineno}

\usepackage[separate-uncertainty=true]{siunitx}
\DeclareSIUnit \UCN{UCN}
\usepackage{multirow}

\begin{document}


\title{First ultracold neutrons produced at TRIUMF}

\newcommand{\UofM}{\affiliation{University of Manitoba, Winnipeg, MB, Canada}}
\newcommand{\UBC}{\affiliation{University of British Columbia, Vancouver, BC, Canada}}
\newcommand{\TRIUMF}{\affiliation{TRIUMF, Vancouver, BC, Canada}}
\newcommand{\UofW}{\affiliation{University of Winnipeg, Winnipeg, MB, Canada}}
\newcommand{\RCNP}{\affiliation{Research Center for Nuclear Physics (RCNP), Osaka University, Osaka, Japan}}
\newcommand{\Tsukuba}{\affiliation{KEK, Tsukuba, Japan}}
\newcommand{\Nagoya}{\affiliation{Department of Physics, Nagoya University, Nagoya, Japan}}
\newcommand{\SFU}{\affiliation{Simon Fraser University, Burnaby, BC, Canada}}

\author{S.~Ahmed}
\UofM
\author{E.~Altiere}
\UBC
\author{T.~Andalib}
\UofM
\author{B.~Bell}
\TRIUMF
\affiliation{McGill University, Montreal, QC, Canada}
\author{C.~P.~Bidinosti}
\UofW
\UofM
\author{E.~Cudmore}
\TRIUMF
\affiliation{Carleton University, Ottawa, ON, Canada}
\author{M.~Das}
\UofM
\author{C.~A.~Davis}
\TRIUMF
\author{B.~Franke}
\TRIUMF
\author{M.~Gericke}
\UofM
\author{P.~Giampa}
\TRIUMF
\author{P.~Gnyp}
\TRIUMF
\affiliation{Coburg University, Coburg, Germany}
\author{S.~Hansen-Romu}
\UofM
\UofW
\author{K.~Hatanaka}
\RCNP
\author{T.~Hayamizu}
\UBC
\author{B.~Jamieson}
\UofW
\author{D.~Jones}
\UBC
\author{S.~Kawasaki}
\Tsukuba
\author{T.~Kikawa}
\affiliation{Kyoto University, Kyoto, Japan}
\author{M.~Kitaguchi}
\Nagoya
\affiliation{Kobayashi-Maskawa Institute for the Origin of Particles and the Universe (KMI), Nagoya University, Nagoya, Japan}
\author{W.~Klassen}
\UofM
\UofW
\author{A.~Konaka}
\TRIUMF
\author{E.~Korkmaz}
\affiliation{University of Northern British Columbia, Prince George, BC, Canada}
\author{F.~Kuchler}
\TRIUMF
\author{M.~Lang}
\UofM
\author{L.~Lee}
\TRIUMF
\UofM
\author{T.~Lindner}
\TRIUMF
\UofW
\author{K.~W.~Madison}
\UBC
\author{Y.~Makida}
\Tsukuba
\author{J.~Mammei}
\UofM
\author{R.~Mammei}
\UofW
\UofM
\TRIUMF
\author{J.~W.~Martin}
\UofW
\UofM
\author{R.~Matsumiya}
\TRIUMF
\author{E.~Miller}
\UBC
\author{K.~Mishima}
\affiliation{J-PARC, Tokai, Japan}
\author{T.~Momose}
\UBC
\author{T.~Okamura}
\Tsukuba
\author{S.~Page}
\UofM
\author{R.~Picker}
\TRIUMF
\SFU
\author{E.~Pierre}
\TRIUMF
\RCNP
\author{W.~D.~Ramsay}
\TRIUMF
\author{L.~Rebenitsch}
\UofM
\UofW
\author{F.~Rehm}
\TRIUMF
\affiliation{Coburg University, Coburg, Germany}
\author{W.~Schreyer}
\email{wschreyer@triumf.ca}
\TRIUMF
\author{H.~M.~Shimizu}
\Nagoya
\author{S.~Sidhu}
\SFU
\TRIUMF
\author{A.~Sikora}
\UofW
\author{J.~Smith}
\TRIUMF
\UBC
\author{I.~Tanihata}
\RCNP
\author{B.~Thorsteinson}
\UofW
\author{S.~Vanbergen}
\TRIUMF
\UBC
\author{W.~T.~H.~van~Oers}
\TRIUMF
\UofM
\author{Y.~X.~Watanabe}
\Tsukuba

\collaboration{TUCAN collaboration}
\noaffiliation

\date{\today}

\begin{abstract}
We installed a source for ultracold neutrons at a new, dedicated spallation target at TRIUMF. The source was originally developed in Japan and uses a superfluid-helium converter cooled to \SI{0.9}{\kelvin}. During an extensive test campaign in November 2017, we extracted up to \num{325000} ultracold neutrons after a one-minute irradiation of the target, over three times more than previously achieved with this source. The corresponding ultracold-neutron density in the whole production and guide volume is \SI{5.3}{\per\cubic\centi\metre}. 
The storage lifetime of ultracold neutrons in the source was initially \SI{37}{\second} and dropped to \SI{24}{\second} during the eighteen days of operation. During continuous irradiation of the spallation target, we were able to detect a sustained ultracold-neutron rate of up to \SI{1500}{\per\second}.

Simulations of UCN production, UCN transport, temperature-dependent UCN yield, and temperature-dependent storage lifetime show excellent agreement with the experimental data and confirm that the ultracold-neutron-upscattering rate in superfluid helium is proportional to $T^7$.

\end{abstract}

\maketitle


\section{Introduction}
\label{sec:intro}

Ultracold neutrons (UCNs) with energies of a few hundred nanoelectronvolts can be trapped by material bottles, magnetic fields, and gravity for hundreds of seconds. That makes them an ideal tool to precisely measure fundamental properties of the neutron, e.g.\ its electric dipole moment~\cite{PhysRevD.92.092003,PhysRevC.92.055501}, lifetime~\cite{Pattie627,PhysRevC.97.055503}, decay correlations~\cite{PhysRevC.87.032501}, interaction with gravitational forces~\cite{ABELE2009593c}, and charge~\cite{SIEMENSEN201526}.

However, the precision of such experiments is limited by the low UCN densities that can be delivered by current sources. Typically, less than two dozen UCNs per \si{\cubic\centi\meter} are detected after filling an experiment~\cite{PhysRevC.95.045503,PhysRevC.97.012501}. The UCN source that has been operating the longest, but still is one of the most intense ones, is installed at Institute Laue-Langevin, Grenoble, France. It reflects cold neutrons on moving blades mounted on a ``UCN turbine'', slowing them to ultracold velocities~\cite{STEYERL1986347}. All newer sources rely on a superthermal process: cold neutrons scattering on a cold converter can induce solid-state excitations and lose almost all of their energy~\cite{GOLUB1975133}. The low temperature of the converter suppresses the inverse process of upscattering.

So far, superthermal sources have been realized with two converter materials: solid deuterium at temperatures around \SI{5}{\kelvin} and superfluid helium (He-II) at temperatures below \SI{1}{\kelvin}. Solid deuterium offers a rich spectrum of solid-state excitations, offering a high UCN-production cross section, but also high absorption cross sections~\cite{PhysRevLett.89.272501}. Conversely, superfluid helium has a lower UCN-production cross section, but can have much lower absorption.

Several superthermal sources with deuterium converters are currently operational, at Los Alamos National Laboratory~\cite{PhysRevC.97.012501}, Paul Scherrer Institut~\cite{LAUSS201498} (both using spallation neutron sources), and at University of Mainz~\cite{Kahlenberg2017} (using a reactor neutron source).

A superfluid-helium converter is used at Institut Laue-Langevin~\cite{PhysRevC.90.015501} (using a cold-neutron beam from a reactor source) and has been used at the Research Center for Nuclear Physics~\cite{PhysRevLett.108.134801} (RCNP, using a spallation neutron source). The latter source has been moved to TRIUMF and installed at a new, dedicated spallation neutron source~\cite{BL1U} in 2017.

\section{Production and losses of ultracold neutrons in superfluid helium}

The dispersion relations of free neutrons and of phonons in superfluid helium cross at an energy $E$ of \SI{1}{\milli\electronvolt}, allowing a neutron with that energy to excite a single phonon and lose virtually all of its energy and momentum. Detailed measurements of the scattering function of superfluid helium show that multi-phonon scattering allows the same process at slightly higher energies~\cite{PhysRevC.92.024004,KOROBKINA2002462}. The UCN-production rate $P$ in the superfluid is given by the cold-neutron flux $\Phi(E)$ and the total scattering cross section $\sigma(E)$ given by these processes:
\begin{equation}
P = \int \Phi(E) \sigma(E) dE.
\label{eq:UCNprod}
\end{equation}

The UCN-loss rate $\tau^{-1}$ in the superfluid, defined as the inverse of the storage lifetime $\tau$, is given by the rates of upscattering in superfluid helium $\tau_\mathrm{up}^{-1}$, absorption in helium $\tau_\mathrm{abs}^{-1}$, wall loss $\tau_\mathrm{wall}^{-1}$, and beta decay $\tau_\beta^{-1}$:
\begin{equation}
\tau^{-1} = \tau_\mathrm{wall}^{-1} + \tau_\mathrm{up}^{-1} + \tau_\mathrm{abs}^{-1} + \tau_\beta^{-1}.
\end{equation}

The wall-storage lifetime is determined by the material, cleanliness, and roughness of the walls and is typically on the order of tens to hundreds of seconds.

The upscattering lifetime is strongly dependent on the temperature $T$ of the superfluid and roughly follows 
\begin{equation}
\tau_\mathrm{up}^{-1} \approx B \cdot \left( \frac{T}{\SI{1}{\kelvin}} \right)^{7},
\label{eq:upscattering}
\end{equation}
with $B$ between \SIlist{0.008;0.016}{\per\second}~\cite{PhysRevC.93.025501}.
So, to suppress the upscattering rate to a similar level as the wall-loss rate, the superfluid helium has to be cooled to a temperature around \SI{1}{\kelvin}.

The absorption lifetime is dominated by the high neutron-absorption cross section of $^3$He. In natural helium---with a $^3$He abundance of \num{1e-6}---the absorption lifetime would be less than \SI{100}{\milli\second}. Isotopically purified helium---available with $^3$He abundances below \num{1e-12}~\cite{HENDRY1987131}---can increase the absorption lifetime to several thousand seconds.

The ultimate limit of storage lifetime is given by the lifetime of free neutrons of $\tau_\beta = \SI{880.2+-1.0}{\second}$~\cite{PDG2018}.

\section{Description of the source}

\begin{figure*}
\centering
\includegraphics[width=\textwidth]{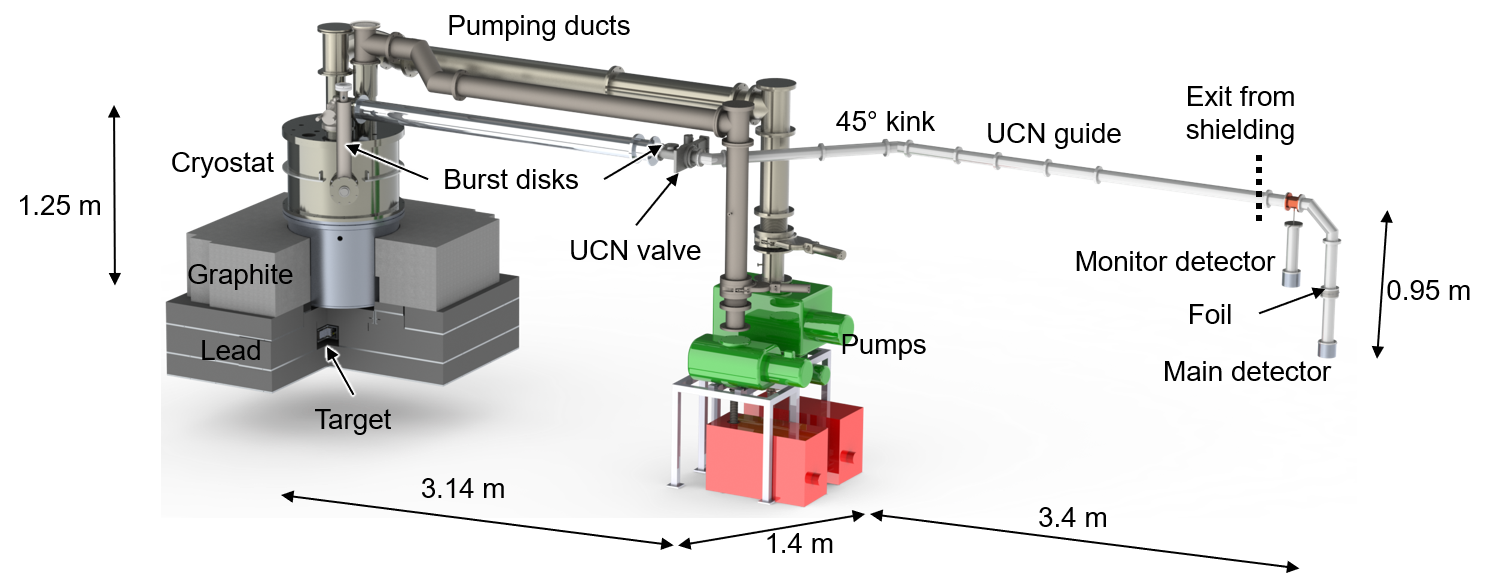}
\caption{UCN source and guide geometry at TRIUMF. When the target is irradiated, spallation neutrons are moderated and converted to ultracold neutrons in the cryostat, see Fig.~\ref{fig:MCNPmodel}. After a period of accumulating UCNs in the source, the UCN valve is opened and UCNs can reach the detectors. The radiation shielding encasing the cryostat and pumps is not shown.}
\label{fig:guidegeometry}
\end{figure*}

\begin{figure}
\centering
\includegraphics[width=0.6\columnwidth]{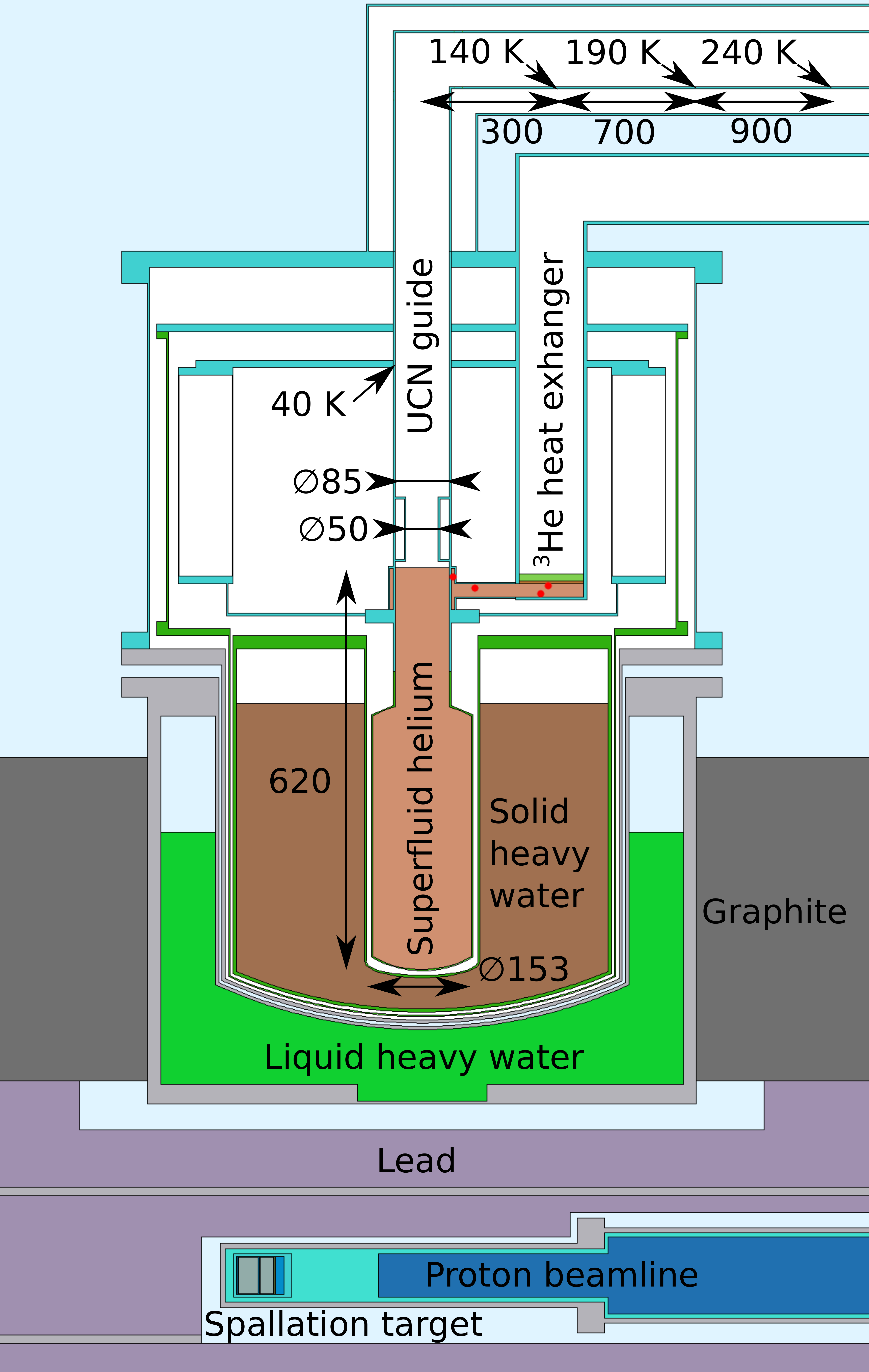}
\caption{Detailed simulation model of the source. Spallation neutrons, produced by irradiating the target with protons, are moderated in heavy water and converted to ultracold neutrons in the superfluid helium. Red dots indicate the temperature sensors used to determine the temperature of the superfluid. The temperature profile on the UCN guide is indicated as well. All dimensions are given in millimeters.}
\label{fig:MCNPmodel}
\end{figure}

The UCN source developed at RCNP uses \SI{8}{\liter} of isotopically purified superfluid helium, cooled to about \SI{0.9}{\kelvin} with a $^3$He cooling circuit. Cold neutrons are provided by two-stage moderation in liquid heavy water at room temperature and solid heavy water cooled to \SI{20}{\kelvin}. For a more detailed description refer to~\cite{PhysRevLett.108.134801}.

In 2017, we moved the source to TRIUMF and installed it at a new, dedicated spallation neutron source. TRIUMF's cyclotron provides a \SI{483}{\mega\electronvolt} proton beam of which up to \SI{40}{\micro\ampere} of beam current can be diverted onto a tungsten spallation target surrounded by lead blocks~\cite{BL1U}. The UCN source is placed above the target and surrounded by graphite blocks serving as additional neutron reflectors. To conform to Canadian safety standards we had to add pressure reliefs on the cryostat and UCN guide and had to add more radiation shielding, requiring a \SI{4.5}{\meter} longer UCN guide than at RCNP, see Fig.~\ref{fig:guidegeometry}.

The UCN-production volume filled with superfluid helium has a cylindrical shape and is attached to a vertical UCN guide, see Fig.~\ref{fig:MCNPmodel}. Heat is conducted from the production volume to a $^3$He-cooled heat exchanger through a single \SI{2}{\milli\meter}-diameter hole in the guide wall and a \SI{0.05}{\milli\meter}-wide gap along its circumference. The temperature of the superfluid helium is measured by four Cernox sensors placed in the superfluid between the UCN guide and the heat exchanger.

The combined height of UCN-production volume and vertical UCN guide is \SI{1.25}{\meter}, with the lower \SI{0.62}{\meter} filled with superfluid helium.  Right above the liquid surface, a short, narrower section of the vertical guide blocks superfluid film flow to reduce heat load. Above the cryostat, the UCN guide continues horizontally in a vacuum jacket to transition from cryogenic to room temperature. It ends with a burst disk for pressure relief and a gate valve (VAT 17.2 series) with a protective ring improving UCN transmission in the open state.

Downstream of the valve, the UCN guide follows a horizontal \SI{45}{\degree} kink to avoid radiation leaking through a direct line of sight to the experimental area. Finally, it penetrates through \SI{3}{\meter} of additional shielding and drops down to allow the UCNs to penetrate a \SI{0.1}{\milli\meter}-thick aluminium foil and to enter the main detector. The total volume of the UCN-production volume and UCN guides is \SI{60.8}{\liter}. The foil separates the helium-filled UCN guide from the detector vacuum to reduce contamination of the source. The main detector uses photomultiplier tubes to detect scintillation light produced by UCNs captured in $^6$Li-enriched glass~\cite{Jamieson2017}. A secondary $^3$He proportional counter with its own aluminium window is mounted to a \SI{5}{\milli\meter} pinhole in the guide, see Fig.~\ref{fig:guidegeometry} and serves as a monitor detector for measurements of transmission through additional guides that will be presented in a separate publication.

\section{Ultracold-neutron yield}

A typical measurement of UCN yield starts with an irradiation of the target with a certain proton-beam current and for a certain duration $t_i$, with the UCN valve closed. During this time, UCNs accumulate in the source, reaching a number
\begin{equation}
N = P \tau_1 \left[ 1 - \exp \left( -\frac{t_i}{\tau_1} \right) \right],
\label{eq:accumulation}
\end{equation}
determined by the production rate $P$ and the loss rate in the source $\tau_1^{-1}$. The loss rate
\begin{equation}
\tau_1^{-1} = f_1 \tau_\mathrm{He}^{-1} + (1 - f_1) \tau_\mathrm{vapor}^{-1} + \tau_\mathrm{wall,1}^{-1} + \tau_\beta^{-1}
\label{eq:sourceloss}
\end{equation}
is the sum of losses in liquid helium $f_1 \tau_\mathrm{He}^{-1}$, in helium vapor $(1 - f_1) \tau_\mathrm{vapor}^{-1}$, on the guide walls $\tau_\mathrm{wall,1}^{-1}$, and due to decay $\tau_\beta$. Since the source is only partially filled with superfluid helium, the loss rate is corrected by the fraction of time $f_1$ that detectable UCNs spend in the superfluid. These components are difficult to disentangle in experiment, instead we estimated them in simulation, see section \ref{sec:simcomparison}.

\begin{figure}
\centering
\includegraphics[width=\columnwidth]{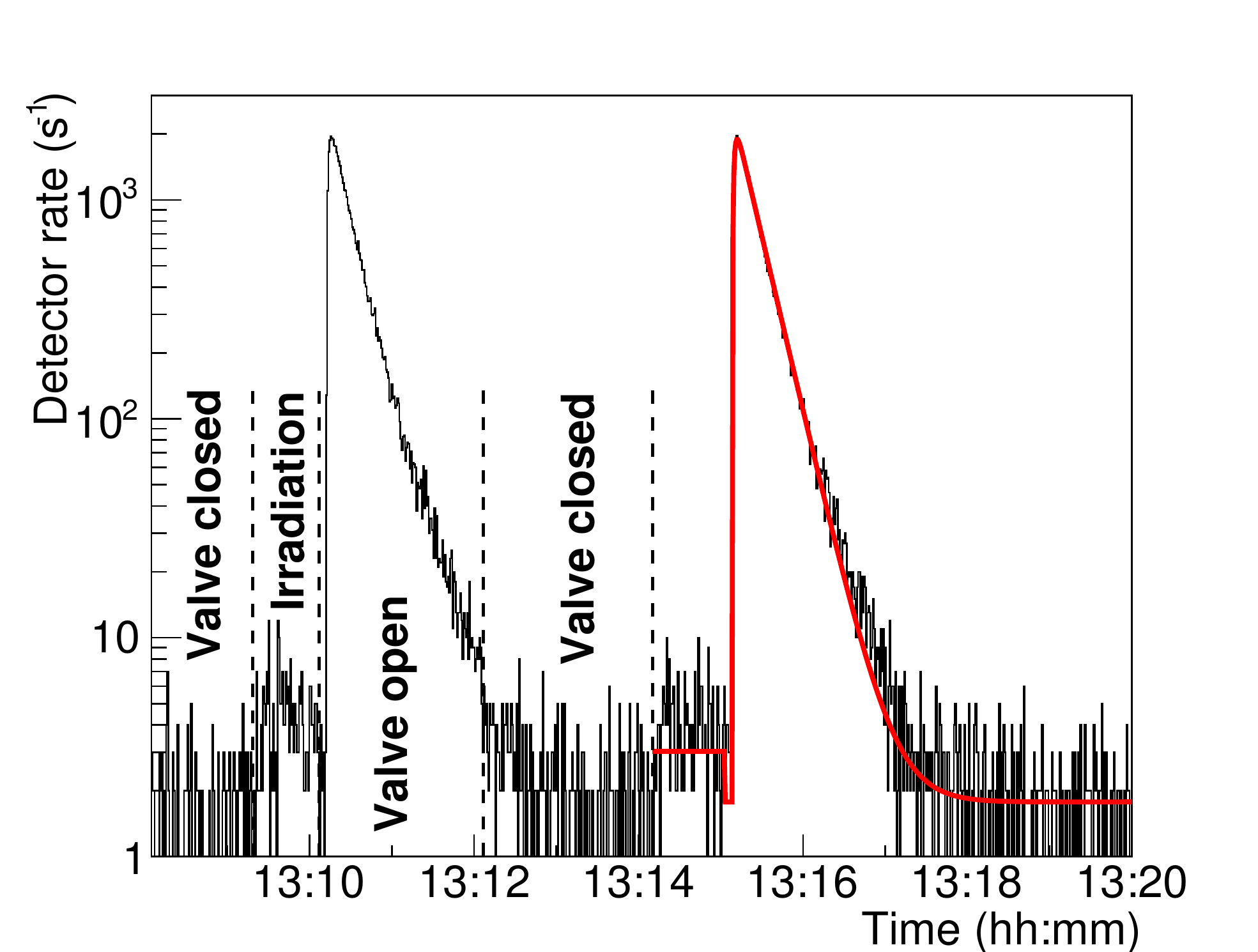}
\caption{Rate in the detector during two typical measurement cycles with a beam current of \SI{1}{\micro\ampere}, an irradiation time of \SI{60}{\second}, and with the valve opened for \SI{120}{\second}. The dashed lines indicate the start of irradiation and the valve actuation times in the first cycle. The red line is a fit as explained in section \ref{sec:simcomparison}.}
\label{fig:rateexample}
\end{figure}

Once the irradiation period ends the valve opens and the accumulated UCNs can reach the detector. The rate in the detector quickly peaks after a few seconds, see Fig.~\ref{fig:rateexample}, and then drops exponentially with a time constant
\begin{equation}
\tau_2^{-1} = f_2 \tau_\mathrm{He}^{-1} + (1 - f_2) \tau_\mathrm{vapor}^{-1} + \tau_\mathrm{wall,2}^{-1} + \tau_d^{-1} + \tau_\beta^{-1}.
\end{equation}
With the valve open, the loss rate to the detector $\tau_d^{-1}$ has to be included. The fraction of time UCNs spend in the superfluid $f_2$ and the wall losses $\tau_\mathrm{wall,2}$ are now different compared to equation (\ref{eq:sourceloss}). The valve stays open for two to three minutes and then the cycle repeats.

We determined the total number of detected UCNs by integrating the rate in the detector while the valve was open and subtracting the background rate, which we estimated before the irradiation started while the valve was closed. During irradiation, the background rate in the detector increased proportionally to the beam current by \SI{2.5+-0.5}{\per\second\per\micro\ampere}. More detailed studies of cross-talk and pile-up in the detector showed that those effects distort the measured rate by less than \SI{1}{\percent}. For details refer to~\cite{Andalib}. To check that the detected neutrons are indeed ultracold neutrons we performed an experiment with  a nickel foil replacing the aluminium foil. In this configuration, the rate in the main detector did not increase above the background, confirming that the vast majority of detected neutrons had energies below \SI{245}{\nano\electronvolt}, the Fermi potential of nickel.

\begin{figure}
\centering
\includegraphics[width=\columnwidth]{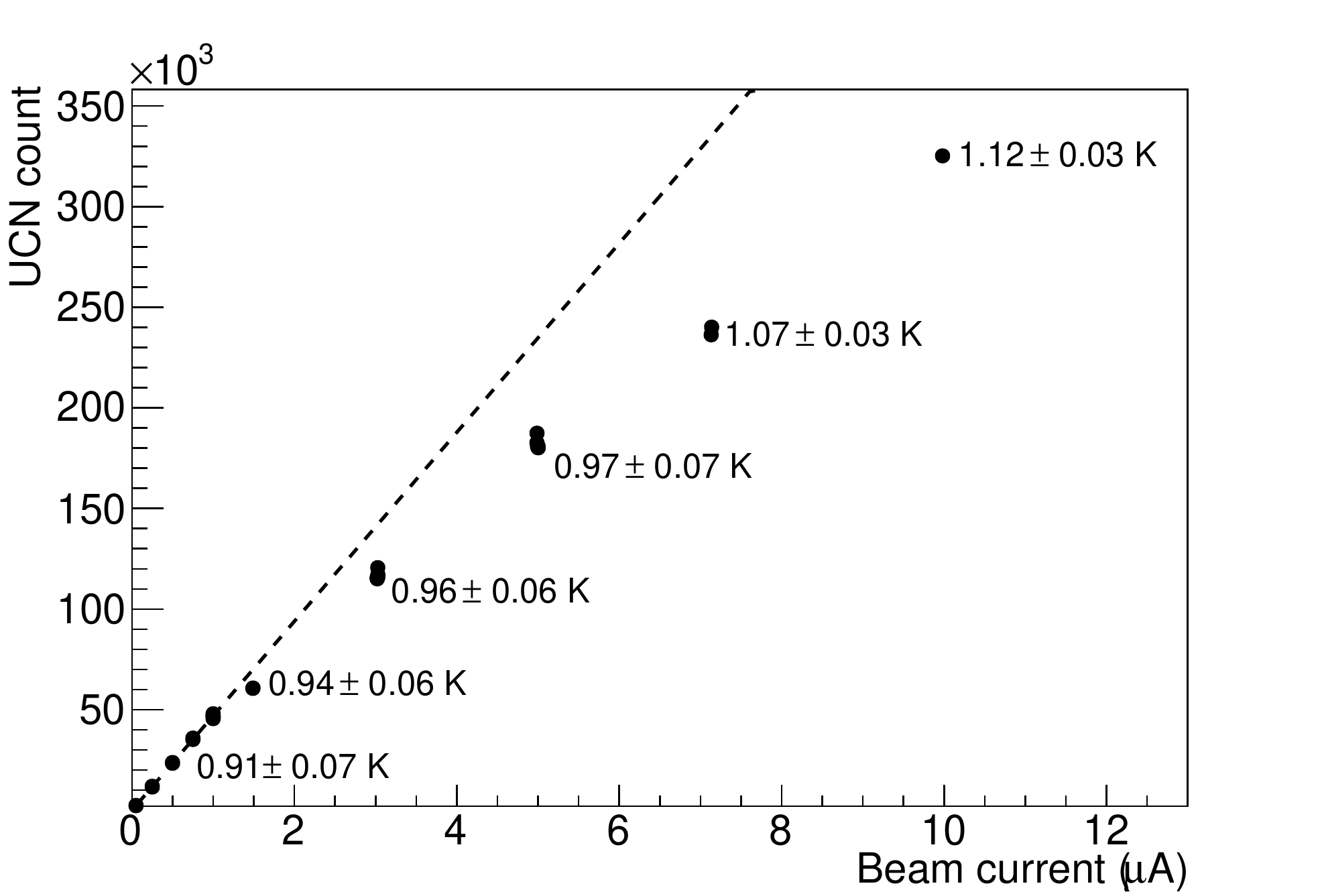}
\caption{Number of UCNs extracted from the source after irradiating the target for \SI{60}{\second} with different beam currents. At currents below \SI{1}{\micro\ampere}, the UCN yield is proportional with current (dashed line). At higher currents, the yield drops due to the increased heat load; the labels indicate the peak helium temperatures reached during irradiation.}
\label{fig:yieldvsbeam}
\end{figure}

The UCN-production rate is expected to be proportional to the beam current. Consequently, for lower beam currents the UCN yield increases linearly with current. However, at higher beam currents the increased heat load on the superfluid raises its temperature and UCN-upscattering rate, reducing the UCN yield, see Fig.~\ref{fig:yieldvsbeam}. The highest number of extracted UCNs was \num{325000} after irradiating the target for \SI{60}{\second} with \SI{10}{\micro\ampere}. Dividing this number by the total guide volume of \SI{60.8}{\liter} yields a UCN density of \SI{5.3}{\per\cubic\centi\meter}. At the nominal beam current of \SI{1}{\micro\ampere} the yield was \num{47500}, corresponding to a UCN density of \SI{0.78}{\per\cubic\centi\meter}.

\begin{figure}
\centering
\includegraphics[width=\columnwidth]{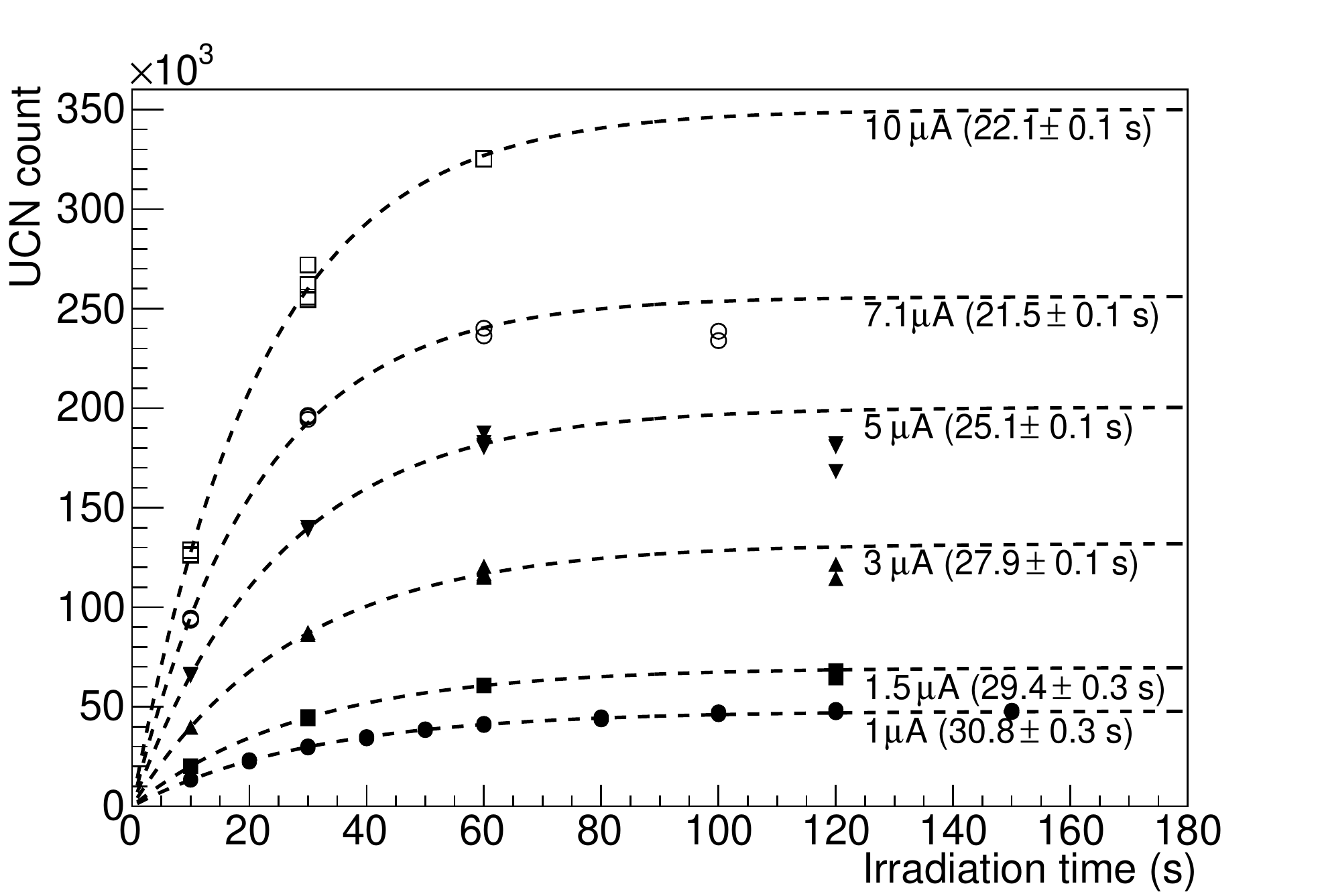}
\caption{Number of UCNs extracted from the source after irradiating the target for various periods with different beam currents. The dashed lines extrapolate the data for irradiation times up to \SI{60}{\second} with exponential saturation curves. The saturation time constant (labels) decreases with higher beam currents.}
\label{fig:yieldvsirradiation}
\end{figure}

The saturating number of UCNs in the source can be directly observed by measuring the UCN yield after different irradiation times, see Fig.~\ref{fig:yieldvsirradiation}. The saturation time constant decreases at higher beam currents, again due to the increasing temperature and upscattering rate of the superfluid. For currents above \SI{1.5}{\micro\ampere} and irradiation times above \SI{60}{\second} the yield starts to drop again due to the further increasing temperature.

Furthermore, instead of operating the source in ``batch mode'', with the valve opening after the irradiation period, we can also continuously irradiate the target while leaving the valve open. At beam currents of \SI{1}{\micro\ampere} or less, such a configuration will lead to a constant stream of \SI{1500}{\UCN\per\second\per\micro\ampere} reaching the detector. During irradiation with higher beam currents, the temperature of the superfluid slowly increases and we observe a decreasing rate.

\section{Storage lifetime}

\begin{figure}
\centering
\includegraphics[width=\columnwidth]{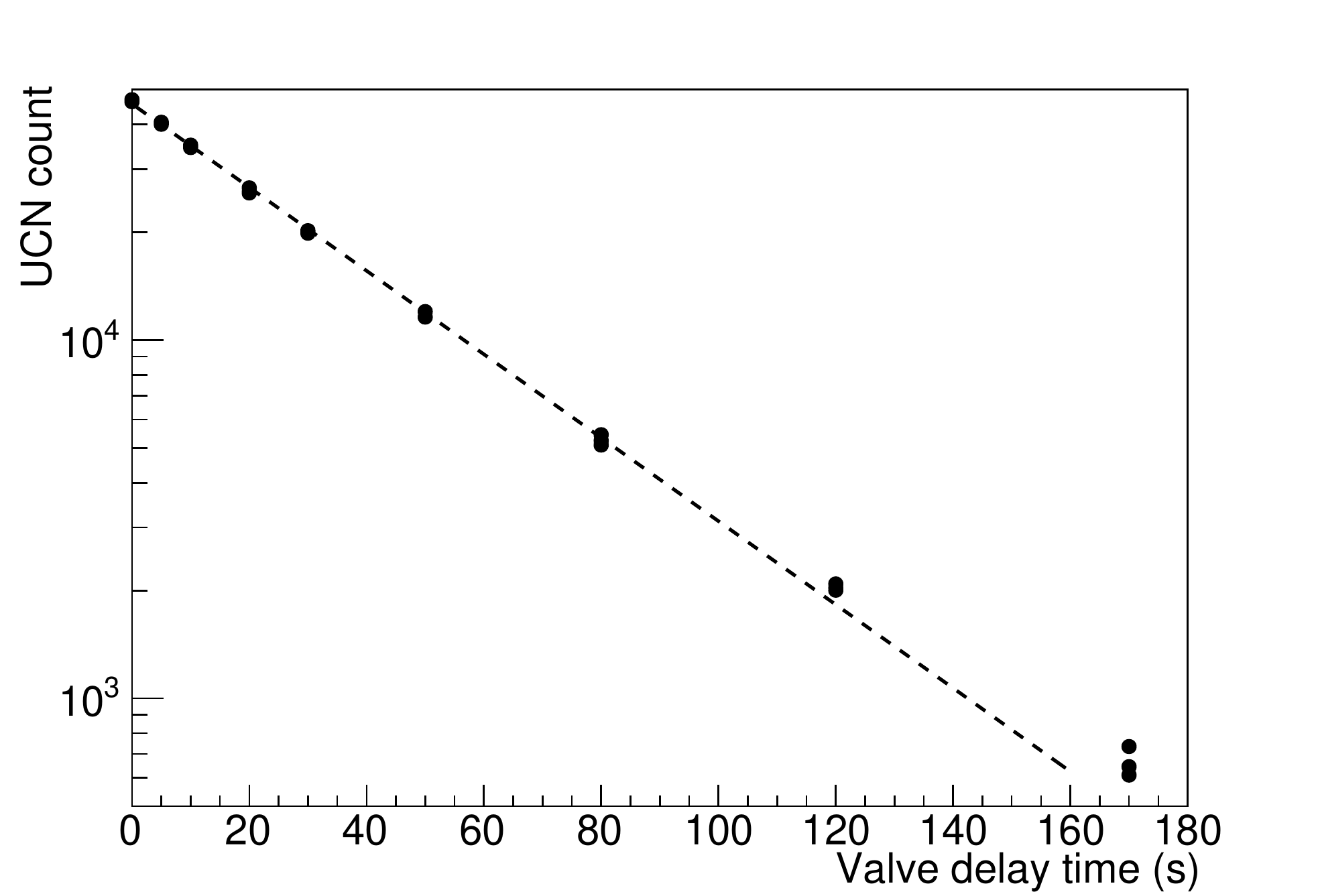}
\caption{UCN yield after irradiating the target for \SI{60}{\second} with \SI{1}{\micro\ampere} while varying the delay between end of irradiation and opening the UCN valve. An exponential fit to the data up to a delay time of \SI{120}{\second} determines the storage lifetime.}
\label{fig:storagelifetimeexample}
\end{figure}

The number of UCNs that can be accumulated directly depends on their storage lifetime in the source $\tau_1$ (equation~\ref{eq:accumulation}), making it a crucial performance parameter. To determine the storage lifetime, we ran cycles where we opened the valve with different delays after the irradiation ended. A typical storage-lifetime measurement consisted of nine cycles with valve delay times of \SIlist{0;170;20;120;50;80;30;20;5}{\second}; an exponential fit through the delay-dependent UCN yield determines the storage lifetime (see Fig.~\ref{fig:storagelifetimeexample}). Although fitting a sum of two exponentials provides a better fit since it takes into account the longer storage lifetimes of low-energy UCNs, we opted for a single exponential fit as the short-term storage lifetimes determine the performance of the source for short irradiation times.

\begin{figure}
\centering
\includegraphics[width=\columnwidth,trim={0 0.7cm 0 0},clip]{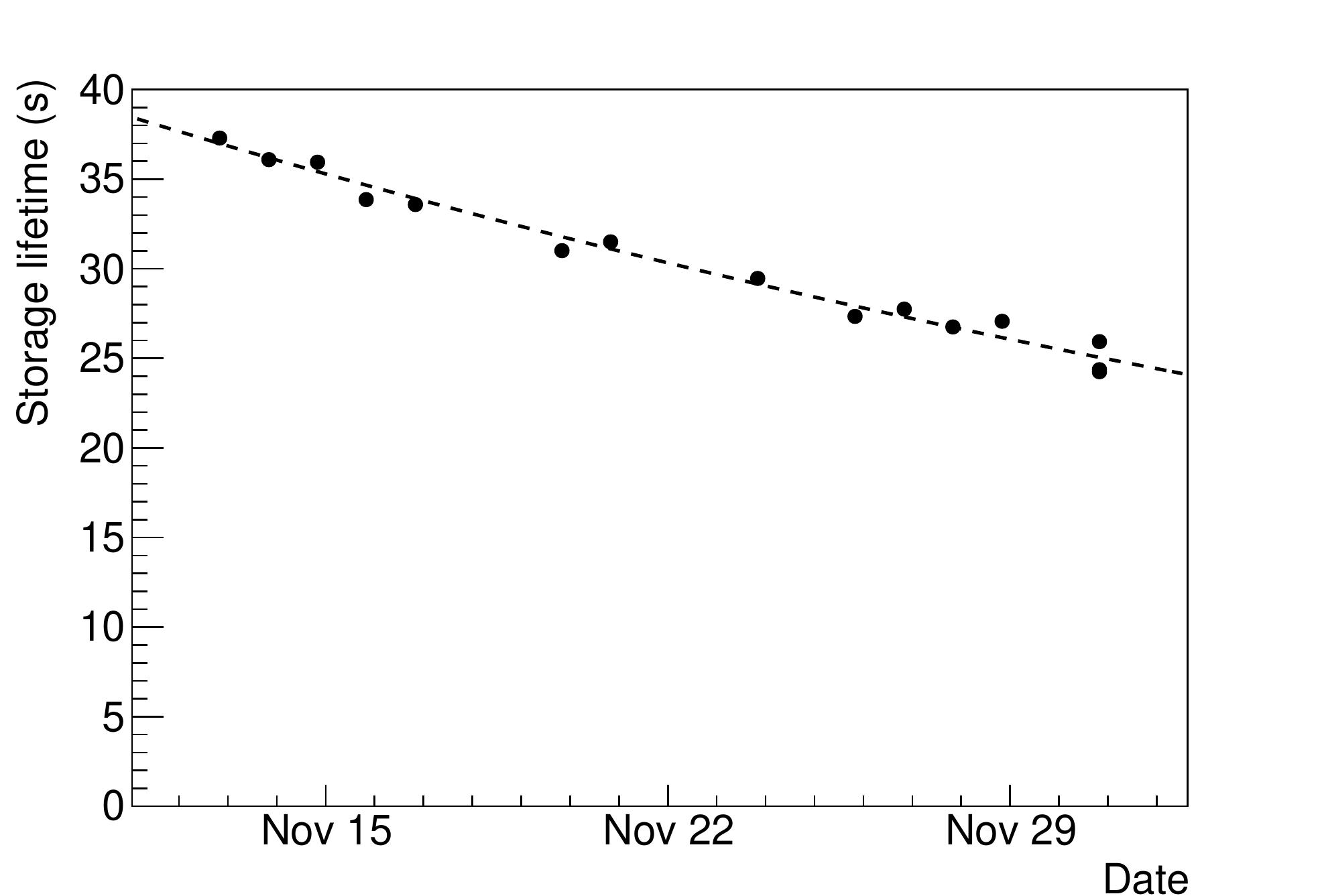}
\caption{Storage lifetime of UCNs in the source on different days after irradiating the target with \SI{1}{\micro\ampere} for \SI{60}{\second}. An exponential fit shows that it dropped by \SI{2.1}{\percent} per day. Uncertainties are smaller than the markers.}
\label{fig:storagelifetimevstime}
\end{figure}

Since the source volume is connected to a long UCN guide sealed with O-rings, we expected residual gas to contaminate the source every time we open the UCN valve. To determine the impact of this contamination, we regularly measured the storage lifetime over a period of eighteen days, see Fig.~\ref{fig:storagelifetimevstime}.
The drop in storage lifetime also directly impacted the UCN yield as expected from equation~\ref{eq:accumulation}.

\section{Comparison with simulations}
\label{sec:simcomparison}

To simulate UCN storage and transport, we built a detailed model of the production volume and UCN guides for the Monte Carlo simulation PENTrack~\cite{SCHREYER2017123}, including the burst disk, actual shape of the UCN valve in open and closed state, pinhole, foil, and main detector. PENTrack uses Fermi potentials to model interaction of UCNs with materials; the imaginary part of the potential determines the loss of UCNs. We set the losses in the foil according to measurements performed by~\cite{ATCHISON2009144}. We modeled the main detector with its two scintillator layers~\cite{Jamieson2017} and their corresponding Fermi potentials and absorption cross sections, as stated in~\cite{Ban2016}.

\begin{table}
\caption{Fermi potentials and diffuse-reflection probabilities used for materials in the PENTrack simulation.}
\begin{ruledtabular}
\begin{tabular}{lrr}
Material & Fermi potential (neV) & Diffusivity \\
\hline
He-II & $18.8 - 0.5 \hbar B T^7 i$ & 0.16 \\
He vapor & $- 0.5 \hbar \tau_\mathrm{vapor}^{-1} i$ & 0 \\
\multirow{2}{*}{Production volume (NiP)} & $213 - 0.100 i$ & \multirow{2}{*}{0.05} \\
                        & $213 - 0.120 i$ & \\
Foil (aluminium) & $54.1 - 0.00281 i$ & 0.20 \\
\multirow{2}{*}{Guides (stainless steel)} & $183 - 0.100 i$ & \multirow{2}{*}{0.03} \\
                        & $183 - 0.140 i$ & \\
GS30 scintillator & $83.1 - 0.000123 i$ & 0.16 \\
GS20 scintillator & $103 - 1.24 i$ & 0.16 \\
\end{tabular}
\end{ruledtabular}
\label{tab:materials}
\end{table}

We assumed that the spectrum of produced UCNs is proportional to $\sqrt{E}$ and that the upscattering rate in superfluid helium follows $\tau_\mathrm{He}^{-1} = B \cdot \left( \frac{T}{\SI{1}{\kelvin}} \right)^7$, with $B$ between \SIlist{0.008;0.016}{\per\second} as measured by~\cite{PhysRevC.93.025501}. By tuning the imaginary Fermi potentials of guides and production volume to \SIrange{0.100}{0.140}{\nano\electronvolt}, see Table~\ref{tab:materials}, we roughly matched the simulated storage lifetime in the source with the measured storage lifetime $\tau_1$. The resulting simulated wall-loss lifetime $\tau_\mathrm{wall,1}$ was \SIrange{32}{38}{\second}.

We also included the upscattering rate in the helium vapor above the liquid $\tau_\mathrm{vapor}^{-1} = \langle v \rangle n \sigma_\mathrm{He}$. This rate depends on the average atomic velocity $\langle v \rangle$ given by the vapor temperature, the vapor density $n$ given by the saturated vapor pressure of the liquid and the vapor temperature, and the thermal-neutron-scattering cross section of $^4$He $\sigma_\mathrm{He} = \SI{0.76}{\barn}$. We assumed that the vapor has the same temperature gradient as measured by several temperature sensors on the outer UCN-guide surface (see Fig.~\ref{fig:MCNPmodel}). To include the temperature gradient in the simulation, we split the guide volume from the liquid surface to the foil into \SI{10}{\centi\meter}-long sections and assigned each an averaged UCN-upscattering rate in this section. Any time-dependent effects, like rapid pressure and temperature changes when the UCN valve is opened and the pressure difference between source and UCN guide is equalized, are not captured with this simple model.

\begin{figure}
\centering
\includegraphics[width=\columnwidth]{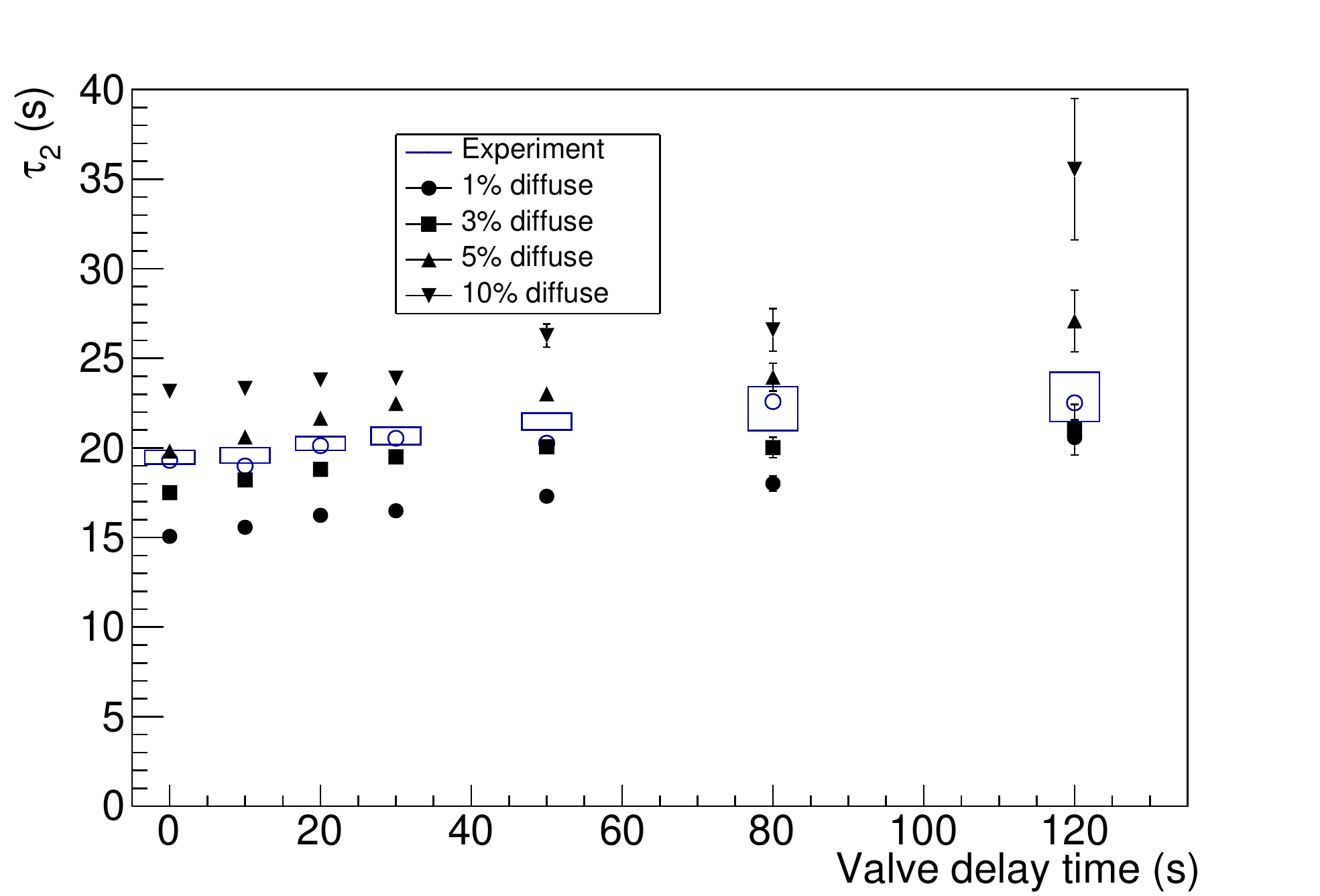}
\includegraphics[width=\columnwidth]{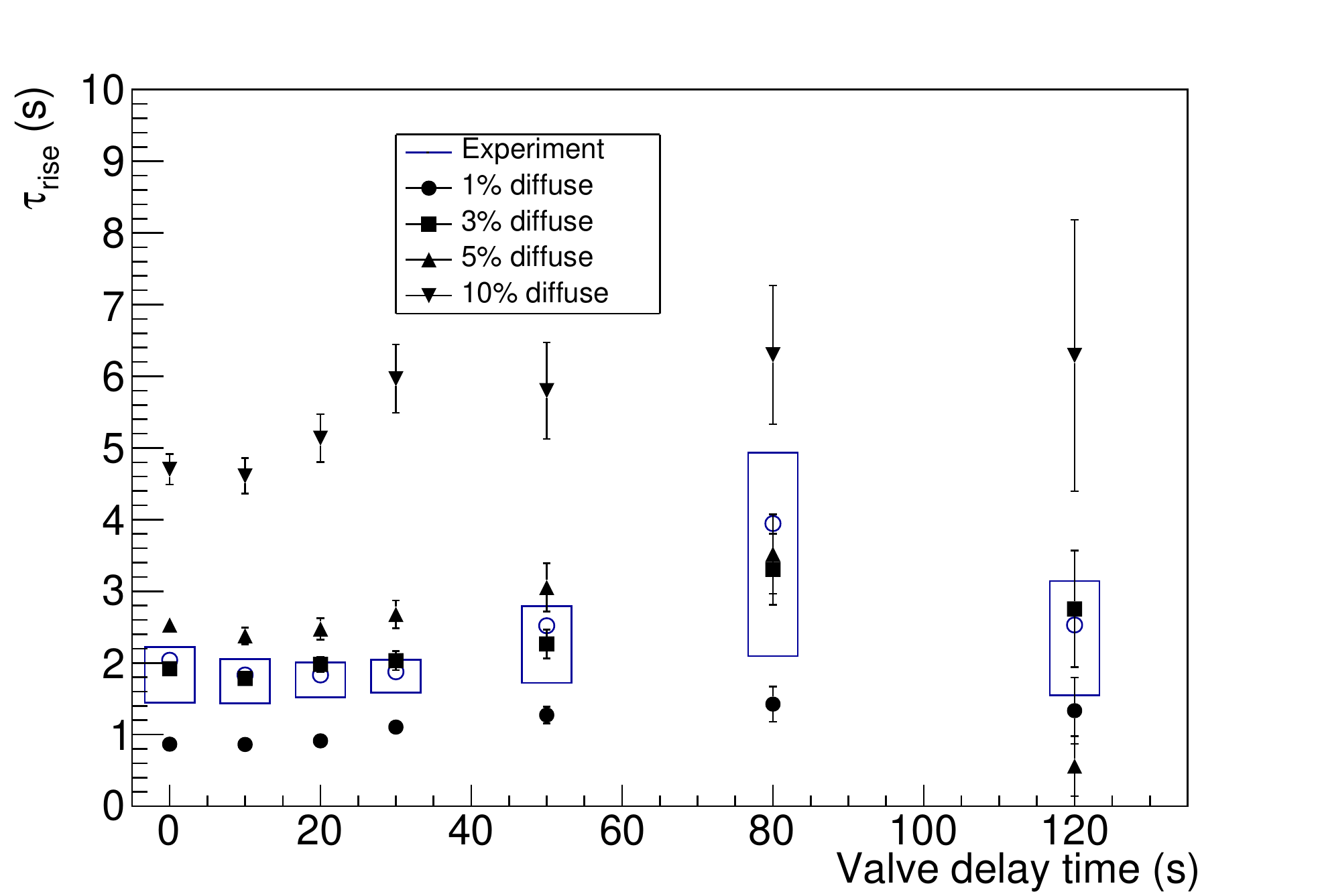}
\caption{Comparison of fall time $\tau_2$ (top) and rise time $\tau_\mathrm{rise}$ (bottom) in experimental data and simulations with different diffuse-reflection probabilities. The boxes indicate the second and third quartile of the experimental data, the empty circle its average. The best match is found with a diffuse-reflection probability of \SI{3}{\percent}.}
\label{fig:falltime}
\end{figure}

To match the simulated UCN transport more accurately to measured data, we fit both the simulated and measured rate of UCNs in the detector after opening the valve at $t = 0$ with the function
\begin{equation}
R(t) = R_0 \left[ 1 - \exp \left( -\frac{t - \Delta t}{\tau_\mathrm{rise}} \right) \right] \exp \left( -\frac{t - \Delta t}{\tau_2} \right) + R_B.
\end{equation}
An example is shown in Fig.~\ref{fig:rateexample}.
Then, we tuned the probability that a UCN is diffusely reflected by the guide walls (following Lambert's law) to match the rise time $\tau_\mathrm{rise}$ and the fall time $\tau_2$ to the experimental data, see Fig.~\ref{fig:falltime}. The delay between opening the valve and the first UCNs being detected in the detector, $\Delta t$, is constant in all scenarios. The parameter $R_B$ is the background rate in the experimental data and zero in the simulated data.

The experimental fall time can be matched with diffuse-reflection probabilities of \SIrange{3}{5}{\percent}, the rise time is best matched with \SI{3}{\percent}. This value is similar to values reported for a range of UCN guides~\cite{DAUM201471,Wlokka:2017wpl,Atchison2010} and we chose it for all subsequent simulations. The time constants slowly change with increasing valve delay times, presumably due to a slow change in the energy spectrum while the UCNs are stored in the source. The simulation also correctly models this behavior.

To estimate UCN production, we built detailed target, moderator, and UCN-converter geometries for the Monte Carlo software MCNP6.1~\cite{doi:10.13182/NT11-135}, taking into account material impurities determined from assays and fill levels of moderator vessels, see Fig.~\ref{fig:MCNPmodel}. With this model, we simulated the complete source: the proton beam hitting the target; secondary neutrons, protons, photons, and electrons; and neutron moderation in graphite and heavy water. In contrast to liquid heavy water, there is no detailed data on thermal-neutron scattering in solid heavy water available. Instead, we relied on a free-gas model with an effective temperature of \SI{80}{\kelvin}, as this seems to be the minimum effective neutron temperature achieved with solid-heavy-water moderators~\cite{doi:10.13182/NSE66-A18558}. From the simulated cold-neutron flux in the UCN-production volume and equation (\ref{eq:UCNprod}) we determined a production rate of \SI{20600+-200}{\per\second\per\micro\ampere} for UCNs with energies up to \SI{233.5}{\nano\electronvolt}.

\begin{figure}
\centering
\includegraphics[width=\columnwidth]{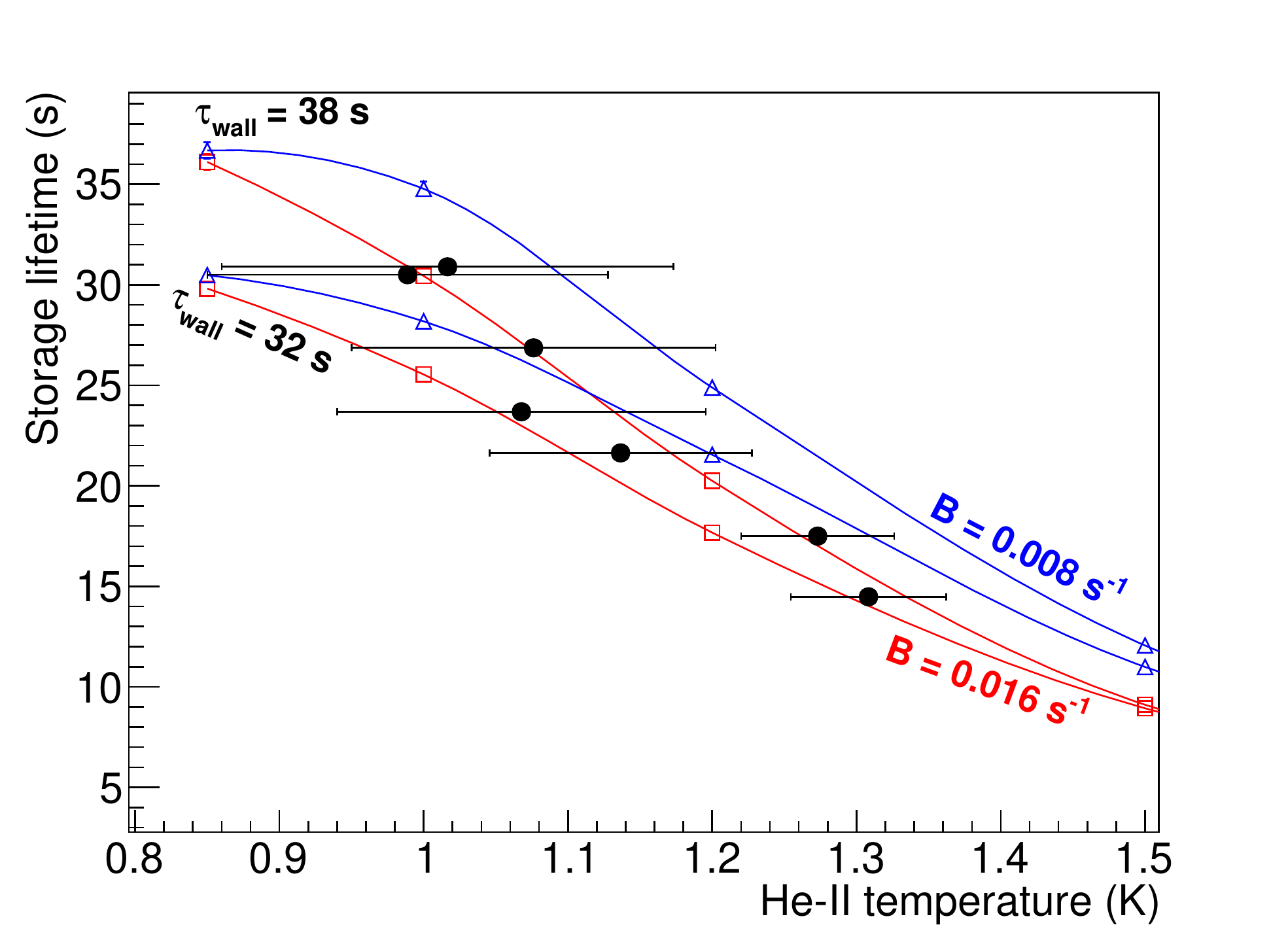}
\includegraphics[width=\columnwidth]{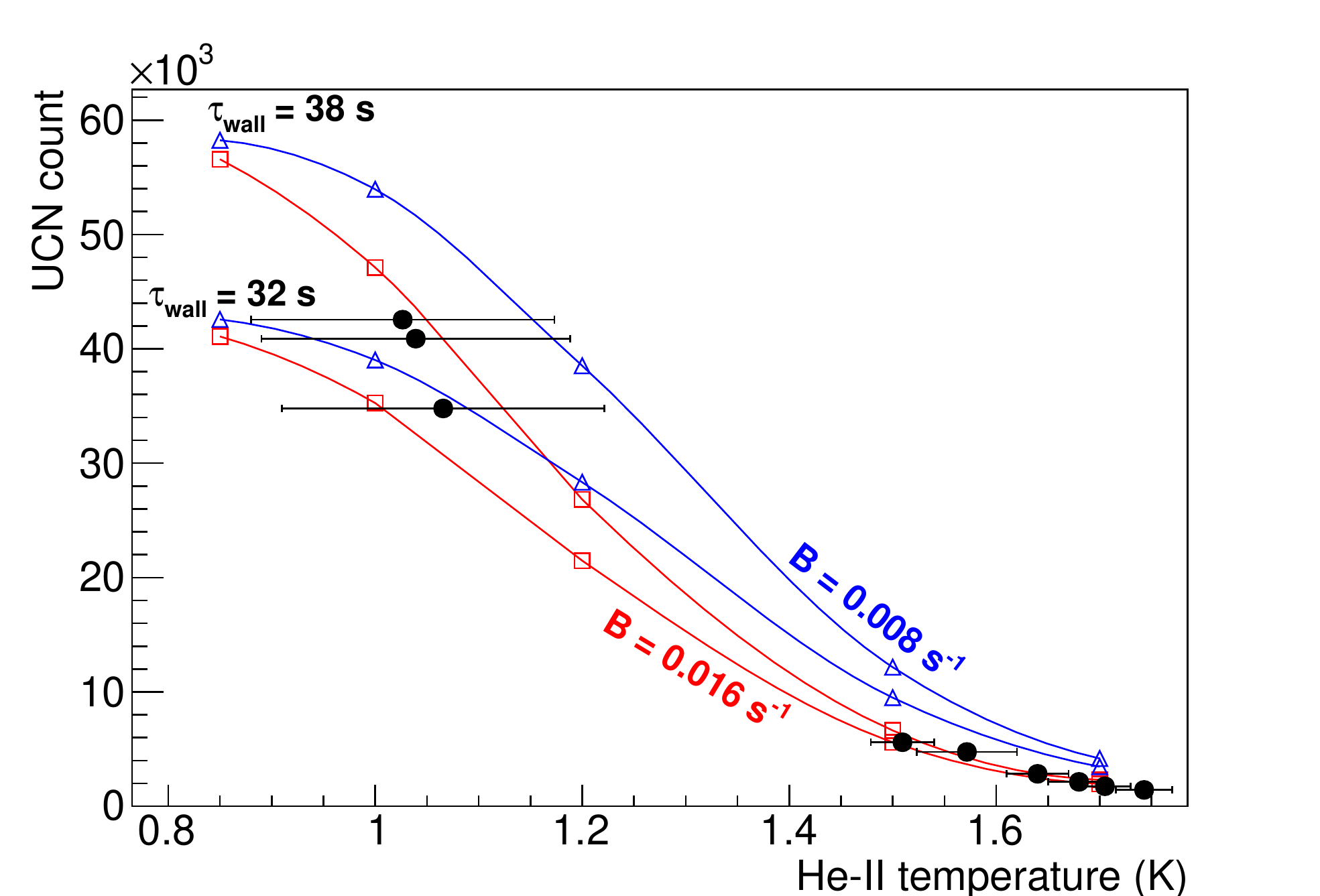}
\caption{Storage lifetime (top) and UCN yield (bottom) at different temperatures after irradiating the target with \SI{1}{\micro\ampere} for \SI{60}{\second} (filled circles). Due to large temperature uncertainties, a range of simulated data (empty squares and triangles) fits the experimental data (see text). The lines are interpolations of simulated data to guide the eye.}
\label{fig:yieldvstemp}
\end{figure}

\begin{figure}
\centering
\includegraphics[width=\columnwidth]{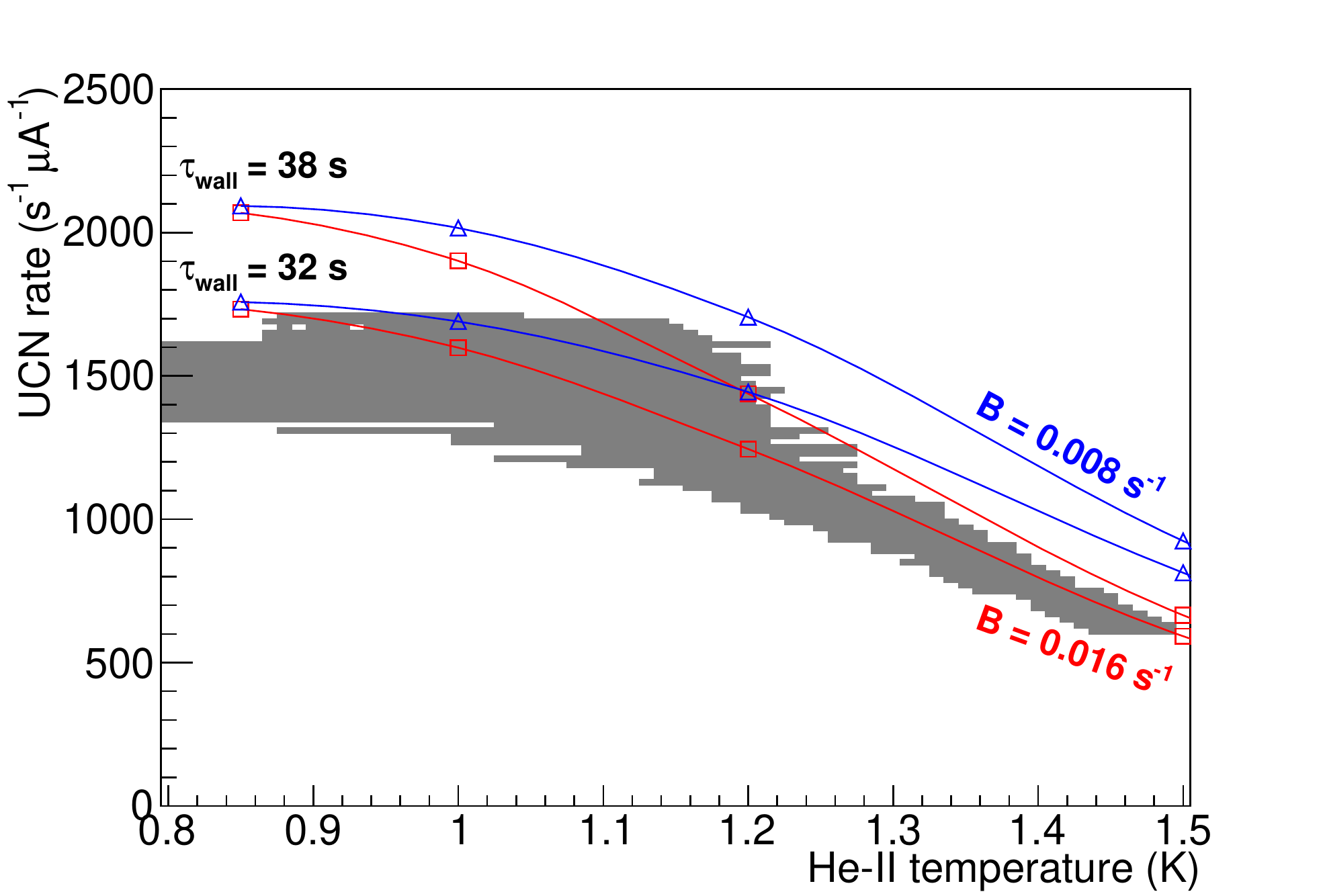}
\caption{Measured UCN rates and temperatures (grey area) while the target is continuously irradiated with the UCN valve open. Due to large temperature uncertainties, a range of simulated data (empty squares and triangles) fits the experimental data (see text). The lines are interpolations of simulated data to guide the eye.}
\label{fig:rate_vs_temp}
\end{figure}

Figure~\ref{fig:yieldvstemp} shows the storage lifetime and yield at different temperatures of the superfluid. In the yield measurement the higher temperatures were reached during an interruption of cooling, in the storage-lifetime measurement by using heaters. Especially at lower temperatures, the four temperature sensors showed large discrepancies. Comparisons with the vapor pressure in the UCN guide, based on the vapor-pressure formula from \cite{PhysRev.100.743}, suggested that the temperature sensors might underestimate the temperature due to poor heat conduction through the small gaps in the guide. Unfortunately, the vapor-pressure measurement was very noisy and had an offset that had to be corrected, increasing the uncertainties at low temperatures even further. The horizontal error bars in Fig.~\ref{fig:yieldvstemp} show the range from the lowest temperature measured by the temperature sensors to the highest temperature derived from both the temperature sensors and the vapor pressure, including the uncertainty in the offset correction, during each measurement.

Due to the large uncertainties, simulations with a range of wall-loss and helium-upscattering parameters fit the data well, see Fig.~\ref{fig:yieldvstemp}. However, when setting $B$ to \SI{0.008}{\per\second} (triangles), the storage lifetime and yield at higher temperatures is slightly overestimated. Simulations without vapor upscattering show significant differences at higher liquid temperatures---at \SI{1.5}{\kelvin} the simulated storage lifetime in vapor, $(1 - f_1)\tau_\mathrm{vapor}$, is reduced to roughly \SI{50}{\second} while it is too large to have a significant effect at \SI{0.9}{\kelvin}.

Figure~\ref{fig:rate_vs_temp} shows the UCN rate in the detector while we continuously irradiated the target with different beam currents with the UCN valve open. At beam currents above \SI{1}{\micro\ampere} the temperature of the superfluid increases slowly, reducing the UCN rate. Again due to the large temperature uncertainties, a range of simulation parameters can match the data, but with $B = \SI{0.008}{\per\second}$ the simulations slightly overestimate the UCN rate at higher temperatures.

Unfortunately, the discrepancies between the temperature sensors and the vapor pressure prevent a more accurate determination of the upscattering parameter. Measurements of heat transport in superfluid helium also agreed with the expected trends in the temperature sensors, once correcting for offsets, and sensor calibration dominated the uncertainty~\cite{Rehm}. We are currently preparing an improved analysis of data with more accurately measured and controlled temperatures and pressures.

\section{Conclusions}

We successfully operated a superfluid-helium source for ultracold neutrons at a new spallation source at TRIUMF. Although we were able to extract three times more UCNs than ever before, thanks to an increased beam current on the spallation target, we achieved only half of the previously best storage lifetime, most likely due to contamination of the source while it was moved, the burst disk added to the UCN guide, and the new UCN valve not optimized for UCN storage.

Simulations including the temperature-dependent upscattering in superfluid helium and helium vapor confirm that the former follows $\tau_\mathrm{He}^{-1} = B \cdot \left( \frac{T}{\SI{1}{\kelvin}} \right)^7$, matching the experimental UCN yield and storage lifetime with $B$ between \SIlist{0.008;0.016}{\per\second}. Upscattering in helium vapor plays a significant role at liquid temperatures above \SI{1}{\kelvin}.

This research provides the prerequisites for future developments: a next-generation source with cooling power and ultracold-neutron flux increased by two orders of magnitude, and an experiment to measure the electric dipole moment of the neutron with a sensitivity of \SI{1e-27}{\elementarycharge\centi\meter}. The excellent match of simulations and experiment makes us confident that we can predict the performance of this future source and experiment very well.

Further operation of the current prototype source will focus on tests of components for these future installations, e.g.\ UCN guides, valves, polarizers, storage volumes, and vacuum windows to mitigate degradation due to contamination.

\begin{acknowledgments}

We would like to thank B.~Hitti and C.~Dick for supplying us with liquid helium; C.~Marshall, T.~Hessels, and S.~Horn for engineering support; C.~Remon, D.~Morris, and the TRIUMF Controls Group for support with controls and DAQ; and the TRIUMF Operations Group for delivering beam to the spallation target.

This work is supported by the Canada Foundation for Innovation (CFI), the Canada Research Chairs program, the Japan Society for the Promotion of Science (JSPS), the Natural Sciences and Engineering Research Council of Canada (NSERC), and Research Manitoba.

Computations were performed on the Cedar supercomputer of WestGrid and ComputeCanada.

\end{acknowledgments}




\bibliography{bibliography}

\end{document}